\documentclass[usegraphicx]{mn2e}

\title[Pumping of OH Main-Line Masers in Star-Forming Regions]
      {Pumping of OH Main-Line Masers in Star-Forming Regions}

\author[M.\,D.\ Gray]
       {M.\,D.\ Gray$^{1}$\\ 
        $^{1}$ School of Physics and Astromomy, University of Manchester,
         Sackville St. Building, PO Box 88, Manchester, M60 1QD, UK}

\begin{document}

\date{Accepted ... .
      Received ... ;
      in original form ...}

\pagerange{000--000}

\maketitle

\begin{abstract}
Pumping routes of masers can in principle be recovered from a
small matrix of master
equations, at an advanced stage of elimination, by tracing back the 
coefficients to a set of unmodified all-process rate coefficients, drawn from
those which appeared in the original set of master equations, prior to any
elimination operations. The traceback is achieved by logging the operations
carried out on each coefficient. There is no guarantee that a pumping scheme
can be represented as a small set of important routes in this way. 
In the present work,
the traceback method is applied to a model which is typical of a large
volume of parameter space which produces very strong inversions in the
main lines of the rotational ground state of OH, at 1665- and 1667-MHz.
For both lines, the pumping scheme is 
largely restricted to the $^{2}\Pi_{3/2}$ 
stack of
rotational levels, and it is possible to list a comparatively small set
of routes (less than ten) which provide more than 80 per cent of the 
inversion. In both cases, the strongest, and simplest, route consists of a
radiative upward stage, to 
the $^{2}\Pi_{3/2}, J=5/2$ rotational level, followed by
a collisional de-excitation to the rotational ground state.

\end{abstract}
\begin{keywords} masers --- molecular processes ---
                 radiation transfer --- ISM: molecules --- radio lines: stars
\end{keywords}

\section{Introduction}

Sophisticated, many-level, non-local-thermodynamic-equilibrium (NLTE)
computations are now routinely used to generate observables, such as
the emergent flux and polarization, from maser regions. Such observables
are based on NLTE molecular energy-level populations and associated radiation fields, which are the fundamental outputs from the calculations. In general,
however, computations of this type yield far less detailed information
about the pumping schemes that produce the maser inversions because the
NLTE solutions do not explicitly store pumping routes. 
The combination of
radiative transfer and kinetic master equations that comprises a model
of a maser environment typically loses track of which pieces of molecular
data, from the (typically) thousands that form the input to the computation,
are responsible for the resulting inversions. 

Our knowledge of maser pumping schemes remains quite patchy. In the case of
OH, the pumping scheme for the \(1612\)-MHz line in long-period variable and
supergiant stars is quite well understood (Elitzur, Goldreich \&
Scoville \shortcite{egs76}; Elitzur \shortcite{el81}). Absorption of \( 35\)-
and \(53\)-\(\mu\)m radiation is required to lift population from the
ground rotational state to the \( J=3/2 \) and \( J=5/2 \) rotational levels
of the \(^{2}\Pi_{1/2} \) stack, followed by a series of radiative decays to the
upper state of the \(1612\)-MHz line. Dickinson \shortcite{dfd87} predicted
from IRAS data that the two pumping lines should make similar contributions
to the pump. A discussion of modern searches for the pumping lines, and
an application of the technique used in the present work (to an OH/IR star
envelope that is far from typical) appears in Gray, Howe \& Lewis 
\shortcite{ghl05}. Investigations into the pumping of the OH ground-state
main lines in stellar envelopes have also been carried out
(Collison \& Nedoluha \shortcite{cn93},\shortcite{cn94}).

A much less complete picture emerges for OH in star-forming regions, 
particularly for the ground-state main lines at \( 1665\) and \(1667\)\,MHz.
One ingredient that is widely believed to be important in the pumping of
these lines is far-infrared (FIR) line overlap (Litvak \shortcite{lit69};
Lucas \shortcite{luc80}; Bujarrabal et al. \shortcite{buj80}). Detailed
models produced after corrected collisional data became available
(Dixon, Field \& Zare \shortcite{dfz85}; Andresen, Hausler \& L\"{u}lf
\shortcite{ahl84}) have added further insights. Piehler \& Kegel
\shortcite{pk89} reject a photodissociative pump, whilst Kylafis \&
Norman \shortcite{kn90} reject a predominantly collisional scheme. Detailed
many-level models, for example, Cesaroni \& Walmsley \shortcite{cw91}, 
Gray, Doel \& Field \shortcite{gdf91}, and
Pavlakis \& Kylafis \shortcite{pk96}
suggest a combination of FIR line overlap and an FIR continuum radiation
field are significant components of the pump, but the details remain obscure.
Recent work on the pumping of \( 1667 \)-MHz masers in megamaser sources
\cite{yu05} links the inversion with FIR radiation at \( 60 \)\,\(\mu\)m.

An alternative approach to the analysis of pumping schemes is required: one
which reveals significantly more detail than broad inferences from NLTE
computations. One such alternative was outlined by Sobolev \shortcite{sob86} and
references thererin. This method involves treating the population flow as
a set of cycles, with varying numbers of links. These cycles were later 
compared by analogy with electronic circuits governed by Kirchoff's Laws
\cite{sobdeg94}. Some subset of these cycles
will be important for sustaining each maser inversion. Finding the strongest
component of the pump was introduced as an analytic optimisation problem,
but the suggestion that was followed is that the whole
method be developed as a Monte-Carlo computer code, providing mainly
statistical information about pumping schemes. This program was used to
analyse the pumping scheme for water masers \cite{sob89}, where it was 
estimated that a very large number of cycles would need to be traced in order
to obtain \(75\) per cent of the population flow.
The most sophisticated use of this Monte-Carlo scheme
is the analysis of pumping routes in methanol masers \cite{sobdeg94,sd94}.
This work was successful in identifying `bottleneck' transitions,
following modifications due to saturation, and in
computing the percentage of the maser flux produced as a function of the
number of links in a cycle. A significant proportion of the flux depended
on complicated cycles involving at least \(25\) links \cite{sobdeg94}.

The method I use to analyse pumping routes in the present work is related
to the flow cycles discussed above \cite{sobdeg94}, but
here the relationship 
between
simple analytic expressions for
the inversion, and rate coefficients, from a partially eliminated 
matrix of master equations, is made explicit. The method also draws
heavily on the simple interpretation of the decomposition of a rate coefficient,
after an arbitrary number of matrix eliminations, into two sub-expressions.
One of these is the same coefficient at an earlier stage of elimination,
and the other is a route via a level equal to the row and column just
eliminated. Further details of the method are explained in Gray, Howe \& Lewis,
\shortcite{ghl05}. A summary is given in Section~2 below.

\section{A summary of the traceback method}

I begin with some definitions. The symbol \( k_{x,y}^{p} \) is an 
all-process rate coefficient, representing the flow of population from
energy level \( x \) to energy level \( y \) in a system of master
equations which is at elimination stage \( p \). For historical reasons, the
elimination stage is defined to be \( p = n+1 \) where \( n \) is the
number of rows (or columns) remaining in the matrix. As elimination proceeds,
\( n \) decreases, and has the value \( 3 \) when the matrix has been
reduced to \( 2 \times 2 \) form. I define the
original matrix to be the matrix formed from the set of master equations
prior to any elimination operations being carried out; it is a square
matrix of size \( N \times N \), where \( N \) is the number of energy
levels in the molecular model. If \( x = y \), the coefficient is diagonal,
and takes on the meaning of the sum of all rate coefficients out of
level \( x \). It can be proved by induction (see Appendix~A) that, provided
this definition of a diagonal coefficient holds at \( p = N+1 \), it holds
for all smaller values of \( n \) until matrix elimination 
reaches a \(2\times 2\) form. Where
it is necessary to raise a coefficient to a power, it is always enclosed in
brackets with the elimination stage inside, and the power outside the brackets.

The coupled radiative transfer and kinetic master equations are solved by
a standard numerical method \cite{jones94}. The same equations are then
re-solved by a naive matrix elimination technique, in which an all-process
rate coefficient, at elimination stage \( p-1 \), can be written as
\begin{equation}
k_{i,j}^{p-1} = k_{i,j}^{p} \pm k_{i,p-1}^{p} k_{p-1,j}^{p} / k_{p-1,p-1}^{p}
\label{coef}
\end{equation}
where the \(-\)sign applies only to modification
of diagonal coefficients, and
where the denominator in the second term on the right-hand side, a diagonal
coefficient, acts as a normalising factor, and this whole term represents the
introduction of a new link to the coefficient, transfering population from
level \( i \) to level \( j \) via level \( p-1 \).

Operations of the type shown in eq.(\ref{coef}) are
logged and divided into three types, depending on the relative sizes of
the two terms on the right-hand side. If the first term is larger than the
second by a factor of at least \( 1/\epsilon \), where \( \epsilon \) is a
parameter less than \( 1 \), the operation is flagged as `unmodified'. This
instructs the {\sc tracer} code, introduced in Gray et al. \shortcite{ghl05},
 to treat \( k_{i,j}^{p-1} \) as unchanged
from its previous value. If the first term is smaller than the second by
a factor smaller than \( \epsilon \), the operation is flagged as 
`replacement': in this case the \( k_{i,j}^{p} \) is discarded in favour of
the new route via level \( p-1 \). The third option (an `amendment')
requires that both parts of the expression be kept.

The computer code {\sc tracer} takes coefficients from a matrix reduced to
a state where \( p \) is small (typically \( 4 \) or \( 5 \)), and expands 
it back to forms with higher \( p \), using the information contained in the
operation log to simplify the expressions as far as possible. Values for
inversions calculated by the standard numerical method are compared with
those computed from coefficients returned by tracer as a check that
\( \epsilon \) has been made small enough (or conversely that sufficient
information has been retained).

Providing that the pumping scheme is sufficiently simple (and there is no
guarantee of this) it is possible to use {\sc tracer} in several stages to
expand coefficients at elimination stage \( p \sim 4 \) back to the original
coefficients of the unmodified matrix, that is where \( p = N+1 \). Once this has been
achieved, the coefficient \( k_{i,j}^{p} \), for small \( p \), has been 
represented in terms of original all-process rate coefficients that can be 
expressed directly in terms of the molecular parameters supplied to the
model (Einstein A-values and collisional rate coefficients) and the radiation
energy density, or mean intensity.

\subsection{inversions}

Expansion of a single coefficient does not, of course, reveal the pumping
scheme. In order to do this, we need to recover the net effect of a set of
coefficients which together yield an inversion in the transition of
interest. The method used is simply to find an analytic expression for the
required inversion in terms of coefficients at small \( p \) and then use
{\sc tracer} on all antagonistic pairs of coefficients which represent
forward (pumping) and reverse (anti-pumping) routes. It is important to
note that some terms, which are large in the {\sc tracer} expansion of
an individual coefficient, may contribute almost nothing to an inversion
because they are paired with a term of the same magnitude in the expansion
of the reverse coefficient. I give below the analytic formulae for the
inversions in the main-line ground state maser transitions of OH. The
formula for \( 1665 \)\,MHz (level \( 3 \) to level \( 1 \)) is given in terms
of coefficients with \( p = 4 \), whilst the analogous formula for
\( 1667 \)\,MHz is written with \( p = 5 \). For a listing of level numbers
in terms of the more usual quantum-mechanical designations for OH, see
Table~\ref{decode}. The level numbers used in the present work are ordered
upwards by energy, with level \( 1 \) being the ground state.
\begin{equation}
\Delta \rho_{3,1} = \frac{{\cal N} k_{2,2}^{4}}{3D} \left\{ \!
   k_{1,3}^{4} - k_{3,1}^{4} + \! \left(
     \frac{k_{1,2}^{4}k_{2,3}^{4}-k_{3,2}^{4}k_{2,1}^{4}}{k_{2,2}^{4}}
                               \! \right) 
                                                  \! \right\}
\label{i1665}
\end{equation}
\begin{eqnarray}
\Delta \rho_{4,2} & \!\!\!\!\! = 
 \!\!\!\!\! & \frac{{\cal N}X}{5Dk_{4,4}^{5}} \left\{\!
   k_{2,4}^{5} - k_{4,2}^{5} +
   \frac{k_{1,1}^{5}}{X}(k_{2,3}^{5}k_{3,4}^{5}-k_{4,3}^{5}k_{3,2}^{5}) \right.
  \nonumber \\
 & \!\!\!\!\! + \!\!\!\!\! &
   \frac{k_{3,3}^{5}}{X}(k_{2,1}^{5}k_{1,4}^{5}-k_{4,1}^{5}k_{1,2}^{5}) \! + \!
   \frac{k_{2,1}^{5}k_{1,3}^{5}k_{3,4}^{5}-
         k_{4,3}^{5}k_{3,1}^{5}k_{1,2}^{5}}{X} \nonumber \\ 
  & \!\!\!\!\! + \!\!\!\!\! & \left.
   \frac{k_{2,3}^{5}k_{3,1}^{5}k_{1,4}^{5}-
         k_{4,1}^{5}k_{1,3}^{5}k_{3,2}^{5}}{X}
                                      \! \right\}
\label{i1667}
\end{eqnarray}
where \( X = k_{1,1}^{5} k_{3,3}^{5} - k_{1,3}^{5} k_{3,1}^{5} \) and
\( \cal N \) is the total number density of OH used in the model.

Equations \ref{i1665} and \ref{i1667} have
 a number of important features: both have,
multiplying the outer bracket, a positive definite 
term which modifies the overall magnitude of the
inversion, but which does not decide its sign. It is obvious that \( X \)
is positive definite because the diagonal term \( k_{1,1}^{5} \) must
contain \( k_{1,3}^{5} \) as part of its sum, and similarly \( k_{3,3}^{5} \)
contains \( k_{3,1}^{5} \). The negative part of the expression, as
written above, is therefore
cancelled exactly. The denominator \( D \) is shown to be positive definite
in Appendix~B. Both inversions are quoted per magnetic sublevel, giving rise
to the \( 3 \) in the denominator of eq.(\ref{i1665}) amd the \(5\), in
eq.(\ref{i1667}).
Inside the outer bracket, groups of coefficients are written in antagonistic
pairs, which
represent the flow of population between the pair of levels that form the
transition of interest. The pair of coefficients involving only the upper
and lower levels of the transition of interest will be referred to as forming
the `direct' route (even though, on expansion, many other levels will in
general be involved). Other pairs, which involve one or two other
un-eliminated levels, will be referred to as forming `indirect' routes,
because they involve levels other than those that make up the transition in
question before any expansion has been performed with {\sc tracer}. The reason
for the greater complexity of eq.(\ref{i1667}) is that it is evaluated for
\( p = 5 \) (a \( 4 \times 4 \) matrix) whilst eq.(\ref{i1665}) uses \( p = 4 \).
\begin{table}
\caption{Quantum-mechanical designations of the first twenty
         hyperfine levels of OH in Hund's case (a) notation.}
\begin{tabular}{@{}lr@{}}
\hline
Level Number                    &      Designation     \\
\hline
1                    &   \( ^{2}\Pi_{3/2}, J=3/2, (1-)  \)         \\
2                    &   \( ^{2}\Pi_{3/2}, J=3/2, (2-)  \)         \\
3                    &   \( ^{2}\Pi_{3/2}, J=3/2, (1+)  \)         \\
4                    &   \( ^{2}\Pi_{3/2}, J=3/2, (2+)  \)         \\ 
5                    &   \( ^{2}\Pi_{3/2}, J=5/2, (2+)  \)         \\
6                    &   \( ^{2}\Pi_{3/2}, J=5/2, (3+)  \)         \\
7                    &   \( ^{2}\Pi_{3/2}, J=5/2, (2-)  \)         \\
8                    &   \( ^{2}\Pi_{3/2}, J=5/2, (3-)  \)         \\
9                    &   \( ^{2}\Pi_{1/2}, J=1/2, (0+)  \)         \\
10                   &   \( ^{2}\Pi_{1/2}, J=1/2, (1+)  \)         \\
11                   &   \( ^{2}\Pi_{1/2}, J=1/2, (0-)  \)         \\
12                   &   \( ^{2}\Pi_{1/2}, J=1/2, (1-)  \)         \\
13                   &   \( ^{2}\Pi_{1/2}, J=3/2, (1-)  \)         \\
14                   &   \( ^{2}\Pi_{1/2}, J=3/2, (2-)  \)         \\
15                   &   \( ^{2}\Pi_{1/2}, J=3/2, (1+)  \)         \\
16                   &   \( ^{2}\Pi_{1/2}, J=3/2, (2+)  \)         \\
17                   &   \( ^{2}\Pi_{3/2}, J=7/2, (4-)  \)         \\
18                   &   \( ^{2}\Pi_{3/2}, J=7/2, (3-)  \)         \\
19                   &   \( ^{2}\Pi_{3/2}, J=7/2, (4+)  \)         \\
20                   &   \( ^{2}\Pi_{3/2}, J=7/2, (3+)  \)         \\
\hline
\end{tabular}
\label{decode}
\end{table}

\section{The OH Pumping Model}

The pump-route traces for both main lines were drawn from the same model.
This was a single computation chosen from a large parameter-space search
using a slab-geometry accelerated lambda iteration (ALI) code \cite{sac85}.
This code incorporates far-infrared (FIR) line overlap 
(Jones et al. \shortcite{jones94}; Stift \shortcite{stift}). It has previously
been applied in the study of several OH and H\(_{2}\)O maser environments
\cite{ghl05,g01,yfg97,ran95}. 

\begin{figure}
\vspace{0.3cm}
\includegraphics[width=8cm]{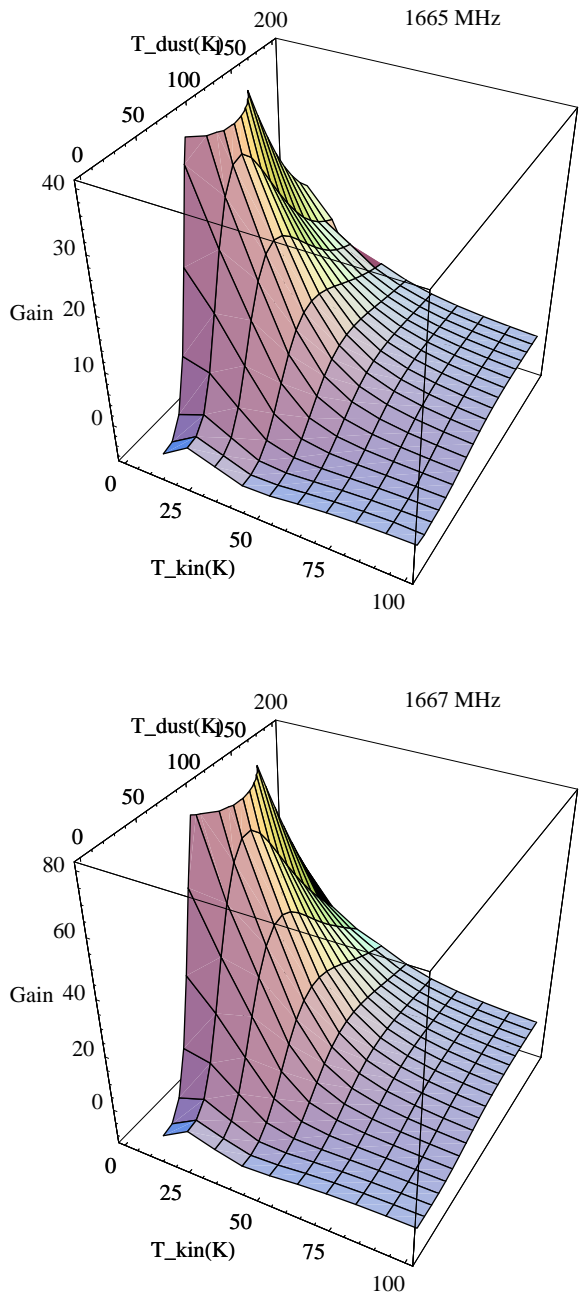}
\caption{
Integrated gains in the \(1665\) (top) and \(1667\)-MHz lines for a
dataset, which includes the chosen
ALI model, plotted against
the kinetic and dust temperatures. Physical conditions not shown on the axes
are those from Table~\ref{parmtab}. Saturation of the maser transitions is
not included.
}
\label{inversionplot}
\end{figure}

The only criterion for selection was that the chosen computation came from
near the peak of the parameter-space search
for unsaturated integrated gain in both the ground-state
main lines. Fig.~\ref{inversionplot} shows unsaturated
 integrated gain plotted as a function
of kinetic and dust temperatures for part of the
parameter space search, including the model chosen. Parameters of the
selected model itself are shown in Table~\ref{parmtab}. Unsaturated
integrated gains predicted for the selected model were \( 20.039 \) at
\( 1665 \)\,MHz and \( 35.661 \) at \( 1667 \)\,MHz. Saturation would limit
any real maser to values in the range \( 10\)-\(14\) as the overall
amplification factor is the exponential of the integrated gain through
the model. Competitive gain in saturation \cite{fg88}, is
effective at converting
initial inversion at \(1667\)\,MHz to maser amplification at \(1665\)\,MHz,
so the model is probably consistent with observations in Galactic star-forming
regions which show that  \(1665\)\,MHz is usually the dominant main-line, and
ground-state, OH maser (for example, Gaume \& Mutel \shortcite{gm87}). An
additional inversion was present in the ground state at \(1720\)\,MHz, but the
only significant excited-state inversions in this model were the
\(^{2}\Pi_{3/2}, J=5/2\) main lines at \(6035\) and \(6030\)\,MHz.

\begin{table}
\caption{Main parameters of the ALI model}
\begin{tabular}{@{}lr@{}}
\hline
Parameter                       &                 Value                   \\
\hline
Total depth                     &   \( 3.0 \times 10^{13} \)\,m           \\
Depth of chosen slab            &   \( 2.0 \times 10^{13} \)\,m           \\
Thickness of chosen slab        &   \( 4.7 \times 10^{12} \)\,m           \\
H\(_{2}\) number density        &   \( 1.0 \times 10^{7}  \)\,cm\(^{-3}\) \\ 
Fractional abundance of OH      &   \( 2.0 \times 10^{-7} \)              \\
Kinetic temperature             &   \( 30 \)\,K                           \\
Dust temperature                &   \( 70 \)\,K                           \\
Bulk velocity shift             &   \( 0.0 \)\,km\,s\(^{-1}\)             \\
Microturbulence                 &   \( 0.0 \)\,km\,s\(^{-1}\)             \\
\hline
\end{tabular}
\label{parmtab}
\end{table}

The tracer method can only work under one set of conditions
at once, including
the radiation field. Therefore, the naive matrix elimination method was run
on a single selected slab drawn from the total of \( 85 \) in the model. These
are arranged logarithmically, such that the depth of slab \( k \) is given
by
\begin{equation}
z_{k} = z_{1} ( z_{M} / z_{1} )^{k/M}
\end{equation}
where $z_{k}$ is the depth of layer $k$, and there are $M$ layers altogether.
The selected slab was \( k = 80 \);
this slab was chosen to have the peak inversion found anywhere in the
model at \( 1665 \)\,MHz. It also turns out that the maximum \( 1667 \)-MHz
inversion was also found in the same slab. The absolute inversions in the
two lines for the chosen slab are \( \Delta \rho_{3,1} = 7.311 \times
10^{-3} \)\,cm\(^{-3}\) and 
\( \Delta \rho_{4,2} = 8.683 \times
10^{-3} \)\,cm\(^{-3}\). The overall number density of OH, from
Table~\ref{parmtab}, is \( 2.0 \)\,cm\(^{-3}\).
The {\sc tracer} analysis which follows, if it
is to have any generality, therefore includes the
assumptions that the pumping schemes in other slabs are generally similar
to those analysed here, and that neighbouring models, such as those
forming the grids in 
Fig.~\ref{inversionplot}, also have pumping schemes broadly similar to
that of the chosen model and slab. This assumption is probably reasonable,
given that the model slab is uniform, and that the integrated gains in
Fig.~\ref{inversionplot} vary smoothly in the vicinity of the chosen model.

\subsection{molecular data}

The molecular data supplied as input to the ALI code comprised energies
of the \( 48 \) hyperfine levels used and Einstein A-values for the
radiatively allowed transitions \cite{Destombes}, complemented by
collisional rate coefficients from
Offer, van Hemmert \& van Dishoeck \shortcite{off94}.
This set of coefficients allows for 
collisions of OH with both ortho- and para-hydrogen. Molecular hydrogen was
assumed to be distributed between the ortho- and para- species at a ratio
of \( 3 \) to \( 1 \).

\section{The 1665\,MHz Pump}

As the highest level involved in the \( 1665 \)-MHz transition is level \( 3\),
it is convenient to begin from a \( 3 \times 3 \) matrix, which yields the
inversion in eq.(\ref{i1665}). For this transition, eq.(\ref{i1665}) has
two antagonistic groups of rate coefficients which control the inversion: the
first is the `direct' route,
\( k_{1,3}^{4} - k_{3,1}^{4}\) and the second, the `indirect' route via
level \( 2 \), 
\( (k_{1,2}^{4} k_{2,3}^{4} - k_{3,2}^{4} k_{2,1}^{4})/k_{2,2}^{4}\). On
examining the magnitudes of these two expressions, the direct route was
found to have the value \(1.05\times 10^{-4}\)\,s\(^{-1}\), and the
indirect route, \(9.07 \times 10^{-5}\)\,s\(^{-1}\). Both terms are positive,
and therefore inverting, and similar in magnitude, so they provide roughly
comparable contributions to the overall pump. It is therefore clear that
both routes must be taken into account when tracing back towards earlier
elimination stages. Fortunately, the trace remains resonably simple, with
two components of the direct route supplying \(83\)\% of it, and
three components of the indirect route contributing \(80.5\)\% of the total in
that system. This total of five dominant terms, which together provide
81.8 per cent of the \( 1665 \)\,MHz inversion, are discussed below in
detail.

\subsection{Route 1}

Route~1 is the strongest term in the expansion of the direct pump,
\( k_{1,3}^{4} - k_{3,1}^{4}\). It has a rate of 
\(6.01 \times 10^{-5}\)\,s\(^{-1}\) (relative strength \(1.0\)) and 
the expression in rate-coefficients
at elimination stage \( 6 \) is
\( (k_{1,5}^{6} k_{5,3}^{6} - k_{3,5}^{6} k_{5,1}^{6})/k_{5,5}^{6}\). Route~1
is the strongest of the five dominant routes,
and also the simplest, because further expansion of the
coefficients in the numerator with {\sc tracer} produces little additional
complexity. Writing unmodified coefficients (at elimination stage \( N+1 \))
without superscript, I found \( k_{1,5}^{6} = k_{1,5} \) and similarly
for \( k_{5,1}^{6} \). The expansion of \( k_{5,3}^{6} \) and \( k_{3,5}^{6} \)
produced original coefficients, and extra terms (via a list of `amendment'
operations). However, all but one of these were found to be weak and
anti-inverting. The remaining route was a weakly inverting pump via
level \( 15 \). However, it was negligible compared to the route using the
original coefficients, and to Route~3 (see below).
Denoting
Route~1 by \( R1 \), the pump route can therefore be expressed as
\begin{equation}
R1 = (k_{1,5} k_{5,3} - k_{3,5} k_{5,1})/k_{5,5}^{6}
\label{R1_1665}
\end{equation}
Note that the denominator in eq.(\ref{R1_1665}) is left evaluated at
elimination stage \( 6 \) for simplicity. As a sum of rate-coefficients, it
is positive definite and cannot control the sign of the expression.
Important original, or unmodified,
all-process rate-coefficients appear 
in Table~\ref{bits1665}. Route~1 is also displayed
by the black arrows in Figure~\ref{pump1665}.

\subsection{Route 2}

Route~2 is the strongest term in the expansion of the indirect pump. It has
the expression,
\begin{equation}
R2 = (k_{1,5}^{6} k_{5,2}^{6} k_{2,4}^{6} k_{4,3}^{6} -
      k_{3,4}^{6} k_{4,2}^{6} k_{2,5}^{6} k_{5,1}^{6})/
             (k_{2,2}^{4} k_{4,4}^{5} k_{5,5}^{6})
\label{R2_1665a}
\end{equation}
and a rate equal to \( 4.86 \times 10^{-5}\)\,s\(^{-1}\).
It has a relative strength (compared to Route~1) of \( 0.808 \).
Unlike Route~1, the expansion of Route~2 with {\sc tracer} 
introduces a complicated
web of routes, not all of which have been traced fully. Three terms control
\( 73 \) per cent of Route~2. When fully expanded back to unmodified
coefficients, these three components of Route~2 are
\begin{eqnarray}
R2 & \!\!\!\!\!\! = \!\!\!\!\!\! & 
\frac{k_{1,5} k_{5,2} k_{2,6} k_{6,4} k_{4,7} k_{7,3} -
               k_{3,7} k_{7,4} k_{4,6} k_{6,2} k_{2,5} k_{5,1}}
              {k_{2,2}^{4} k_{4,4}^{5} k_{5,5}^{6} k_{6,6}^{5} k_{7,7}^{8}}
              \nonumber \\
   & \!\!\!\!\!\! + \!\!\!\!\!\! &
\frac{k_{1,5} k_{5,2} k_{2,6} k_{6,4} k_{4,8} k_{8,20} k_{20,7} k_{7,3}-
              \Omega_{2B}}
             {k_{2,2}^{4} k_{4,4}^{5} k_{5,5}^{6} k_{6,6}^{5} k_{7,7}^{8}
               k_{8,8}^{9} k_{20,20}^{21}} \nonumber \\
   & \!\!\!\!\!\! + \!\!\!\!\!\! &
\frac{k_{1,5} k_{5,2} k_{2,10} k_{10,14} k_{14,4} k_{4,7} k_{7,3}-
              \Omega_{2C}}
             {k_{2,2}^{4} k_{4,4}^{5} k_{5,5}^{6} k_{10,10}^{5} 
              k_{14,14}^{15} k_{7,7}^{8}}
\label{R2_1665}
\end{eqnarray}
where \( \Omega_{2B} \) and \( \Omega_{2C} \) are reverse routes, formed from
a product of rate coefficients with each pair of levels reversed with respect
to those of the forward route; \( \Omega_{2A}\) is written out in full. The
lines of eq.(\ref{R2_1665}) correspond to the A, B and C subroutes in
Fig.~\ref{pump1665}.
The loss of \( 27 \) per cent of Route~2, missing from eq.(\ref{R2_1665}),
arises from ignoring \( k_{5,6}^{7} k_{6,2} / k_{6,6}^{7} \) in favour of
\( k_{5,2} \) in the expansion of \( k_{5,2}^{6} \) (and a similar 
approximation in the reverse route). This omission is comparable, in
magnitude, to ignoring the fourth and fifth dominant terms (the remaining two
from the indirect pump). The second and third lines in eq.(\ref{R2_1665}) are
of comparable strength, and these are both about 1/5 of the strength of the
route on the first line. The web of transitions comprising Route~2 is
represented by blue arrows in
Fig.~\ref{pump1665} (except for transitions already marked
as part of Route~1).

\subsection{Route 3}

Route~3 comes from the direct part of the pump, and has a rate of
\(2.72 \times 10^{-5}\)\,s\(^{-1}\) (relative strength \(0.453\)), placing it
third in strength of the five dominant terms. The expression for \( R3 \)
is
\begin{equation}
(k_{1,4}^{5} k_{4,3}^{5} - k_{3,4}^{5} k_{4,1}^{5})/k_{4,4}^{5}
\label{R3_1665a}
\end{equation}
The expansion of Route~3 leads to a complex web of routes most of which, for
brevity, have been omitted here. The strongest three routes, when fully
traced back to unmodified coefficients, appear in eq.(\ref{R3_1665}). These
three together account for only \( 50 \) per cent of Route~3. Six terms
appear in eq.(\ref{R3_1665}) because some of the expansion involves
amendment operations which add terms. The first four terms correspond to
Route~3A, and the last pair are Route~3B and Route~3C, respectively.
Reverse routes are written in full
except for in the third line. 
\begin{eqnarray}
R3 & \!\!\!\!\!\! = \!\!\!\!\!\! & 
\frac{k_{1,9} k_{9,4} k_{4,7} k_{7,3} -
               k_{3,7} k_{7,4} k_{4,9} k_{9,1}}
              {k_{4,4}^{5} k_{7,7}^{8} k_{9,9}^{10}}
              \nonumber \\
   & \!\!\!\!\!\! + \!\!\!\!\!\! &
\frac{k_{1,9} k_{9,4} k_{4,8} k_{8,20} k_{20,7} k_{7,3}-
              k_{3,7} k_{7,20} k_{20,8} k_{8,4} k_{4,9} k_{9,1}}
             {k_{4,4}^{5} k_{7,7}^{8} k_{8,8}^{9}
                k_{9,9}^{10} k_{20,20}^{21}} \nonumber \\
   & \!\!\!\!\!\! + \!\!\!\!\!\! &
\frac{k_{1,9} k_{9,13} k_{13,10} k_{10,14} k_{14,4} k_{4,7} k_{7,3}-
              \Omega_{3Aiii}}
             {k_{4,4}^{5} k_{7,7}^{8} k_{9,9}^{10} k_{10,10}^{11}
               k_{13,13}^{14} k_{14,14}^{15}} \nonumber \\
   & \!\!\!\!\!\! + \!\!\!\!\!\! &
\frac{k_{1,9} k_{9,13} k_{13,4} k_{4,7} k_{7,3}-
              k_{3,7} k_{7,4} k_{4,13} k_{13,9} k_{9,1}}
             {k_{4,4}^{5} k_{7,7}^{8} k_{9,9}^{10} k_{13,13}^{14}} \nonumber \\
   & \!\!\!\!\!\! + \!\!\!\!\!\! &
\frac{k_{1,10} k_{10,14} k_{14,4} k_{4,7} k_{7,3}-
              k_{3,7} k_{7,4} k_{4,14} k_{14,10} k_{10,1}}
             {k_{4,4}^{5} k_{7,7}^{8} 
              k_{10,10}^{11} k_{14,14}^{15}} \nonumber \\
   & \!\!\!\!\!\! + \!\!\!\!\!\! & 
\frac{k_{1,5} k_{5,4} k_{4,3} -
               k_{3,4} k_{4,5} k_{5,1}}
              {k_{4,4}^{5} k_{5,5}^{6}}
\label{R3_1665}
\end{eqnarray} 

\subsection{Route~4}

Route~4 comes from the indirect part of the pump; it has a rate equal to
\begin{equation}
R4 = (k_{1,5} k_{5,2} k_{2,6} k_{6,3}^{7} -
      k_{3,6}^{7} k_{6,2} k_{2,5} k_{5,1})/
             (k_{2,2}^{4} k_{5,5}^{6} k_{6,6}^{7})
\label{R4_1665}
\end{equation}
with a numerical value of \(1.31 \times 10^{-5}\)\,s\(^{-1}\) (relative
strength = \( 0.218 \)).

\subsection{Route~5}

Route~5 is also part of the indirect pump; it has a rate equal to
\begin{equation}
R5 = \frac{k_{1,5} k_{5,2} k_{2,10} k_{10,3}^{11} -
      k_{3,10}^{11} k_{10,2} k_{2,5} k_{5,1}}
             {k_{2,2}^{4} k_{5,5}^{6} k_{10,10}^{11}}
\label{R5_1665}
\end{equation}
with a numerical value of \(1.13 \times 10^{-5}\)\,s\(^{-1}\) (relative
strength = \( 0.188 \)).

\begin{figure}
\vspace{0.3cm}
\includegraphics[width=8cm]{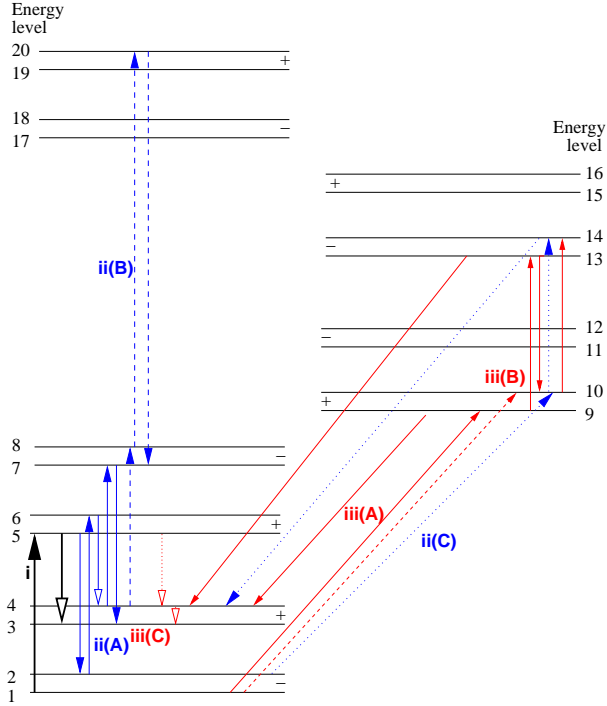}
\caption{
Principal pumping routes for the \( 1665 \)\,MHz maser under the conditions of
Table~\ref{parmtab}. Energy levels are numbered following the scheme in
Table~\ref{decode}, and are marked with a parity assignment (+ or -). 
Energy gaps between levels are not drawn to scale. Route~1
transitions are shown in black; additional transitions appearing in Route~2
are drawn in blue, and further additions from Route~3 are in red. Transitions
appearing from an `A' sub-route have solid lines, those from a `B' sub-route
are marked by dashed lines, those from a `C' subroute, by dotted lines.
For
simplicity, only the forward (pumping) routes are shown for each antagonistic
pair, and transitions appearing in more than one route are drawn only
once. A solid arrowhead indicates a transition allowed for electric dipole
radiation; a hollow arrowhead indicates a radiatively forbidden transition
in which population transfer can be considered to proceed via collisions only.
}
\label{pump1665}
\end{figure}

\subsection{Summary}

The \( 1665 \)\,MHz pump, though complex, is simple enough to allow a
large percentage of its strength to be represented in a modest number
of terms. The overall degree of complexity is similar to that found for
the \( 1612 \)\,MHz line in the OH/IR stellar envelopes studies in Gray, Howe
\& Lewis \shortcite{ghl05}. In contrast to that case, in which transfer
to the \( ^{2}\Pi_{1/2} \) stack from the ground state by \( 53 \)\,\(\mu\)m radiation
is amongst the most important processes, the \( 1665 \)\,MHz pump tends to
stay largely within the \( ^{2}\Pi_{3/2} \) stack, with transfer to \( ^{2}\Pi_{1/2} \) appearing
only in weaker terms of Route~2, and in Route~3. The strongest pump route
of all, Route~1, is extremely simple, comprising only two parts: a radiative
FIR absorption from the ground state up to \( ^{2}\Pi_{3/2}, J=5/2 \), and a collisional
decay back to level \( 3 \) in the upper part of the ground-state lambda
doublet. This downward leg must be collisional because the upper part of
the ground-state lambda doublet and the lower part of the excited-state
lambda doublet have the same parity, making this transition radiatively
forbidden.

Most of the transitions shown in Fig.~\ref{pump1665} are radiative, but there
are a significant number of important links which must be predominantly 
collisional, either because the transitions involved are within
lambda doublets and have a very small A-value, or they are radiatively
forbidden by parity or other quantum-mechanical selection rules. No part
of the pumping scheme involves levels higher than 
those in the \( ^{2}\Pi_{3/2}, J=7/2 \)
lambda doublet.

\section{The 1667\,MHz Pump}

\begin{figure}
\vspace{0.3cm}
\includegraphics[width=8cm]{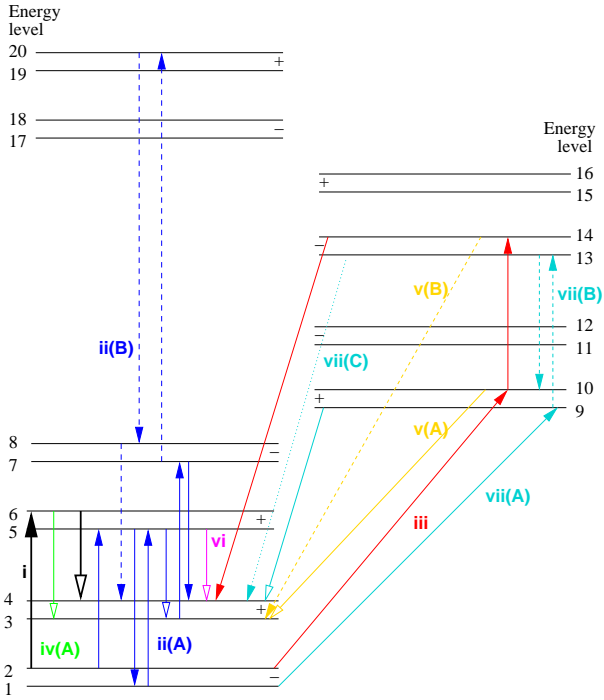}
\caption{
Principal pumping routes for the \( 1667 \)\,MHz maser under the conditions of
Table~\ref{parmtab}. Markings as for Fig.~\ref{pump1665} except that the
colour code is extended to seven colours: transitions appearing in Route~4
are drawn in green and then Route~5 in gold, Route~6 in pink and Route~7
in turquoise.
}
\label{f:fig1667}
\end{figure}

This pumping scheme was traced from a \( 4 \times 4 \) matrix, with the
inversion given by eq.(\ref{i1667}). The inversion expression is therefore
considerably more complicated than in the case of \( 1665 \)\,MHz, but the
compensation for this is reduced complexity in the traceback. Seven
contributions to the inversion were found which had an inverting effect of
at least ten per cent of the strongest. Of these seven routes, the
strongest pair account for \( 56.5 \) per cent of the total inversion. In
fully traced-back form, the seven routes become:
\begin{equation}
R1 = \frac{k_{2,6} k_{6,4} - k_{4,6} k_{6,2}}{k_{6,6}^{7}}
\label{R1_1667}
\end{equation}
where \( R1 \) is taken to have a relative strength of \( 1.0 \), and
\begin{eqnarray}
R2 & \!\!\!\!\! = \!\!\!\!\! & 
\frac{k_{2,5} k_{5,1} k_{1,5} k_{5,3} k_{3,7} k_{7,4} -
          k_{4,7} k_{7,3} k_{3,5} k_{5,1} k_{1,5} k_{5,2}}
         {[k_{5,5}^{6}]^{2} k_{7,7}^{8} X} \nonumber \\
   & \!\!\!\!\! + \!\!\!\!\! &
\frac{k_{2,5} k_{5,1} k_{1,5} k_{5,3} k_{3,7} k_{7,20} k_{20,8} k_{8,4}-
              \Omega_{R2}}
         {[k_{5,5}^{6}]^{2} k_{7,7}^{8} k_{8,8}^{9} k_{20,20}^{21} X} =
         0.413 R1
\label{R2_1667}
\end{eqnarray}
\begin{equation}
R3 = \frac{k_{2,10} k_{10,14} k_{14,4} - k_{4,14} k_{14,10} k_{10,2}}
     {k_{10,10}^{11} k_{14,14}^{15}}
     = 0.290 R1
\label{R3_1667}
\end{equation}
\begin{eqnarray}
R4  & \!\!\!\!\! = \!\!\!\!\! & 
\frac{k_{1,1}^{5} (k_{2,6} k_{6,3} k_{3,7} k_{7,4} -
                             k_{4,7} k_{7,3} k_{3,6} k_{6,2})}
               {k_{7,7}^{8} k_{8,8}^{9} X} \nonumber \\
    & \!\!\!\!\! + \!\!\!\!\! & 
\frac{k_{1,1}^{5} (k_{2,6} k_{6,3} k_{3,7} k_{7,20} k_{20,8} k_{8,4}-
                             \Omega_{R4})}
               {k_{7,7}^{8} [k_{8,8}^{9}]^{2} k_{20,20}^{21} X } =
         0.255 R1
\label{R4_1667}
\end{eqnarray}
\begin{eqnarray}
R5  & \!\!\!\!\! = \!\!\!\!\! & 
\frac{k_{1,1}^{5} (k_{2,10} k_{10,3} k_{3,7} k_{7,4} -
                             k_{4,7} k_{7,3} k_{3,10} k_{10,2})}
               {k_{7,7}^{8} k_{10,10}^{11} X} \nonumber \\
    & \!\!\!\!\! + \!\!\!\!\! & 
\frac{k_{1,1}^{5} (k_{2,10} k_{10,14} k_{14,3} k_{3,7} k_{7,4}-
                             \Omega_{R5})}
               {k_{7,7}^{8} k_{10,10}^{11} k_{14,14}^{15} X} =
         0.228 R1
\label{R5_1667}
\end{eqnarray}
\begin{equation}
R6 = \frac{k_{3,3}^{5} (k_{2,5} k_{5,1} k_{1,5} k_{5,4} \! - 
                      \!  k_{4,5} k_{5,1} k_{1,5} k_{5,2})}
          {[k_{5,5}^{6}]^{2} X} = 0.203 R1
\label{R6_1667}
\end{equation}
\begin{eqnarray}
R7 & \!\!\!\!\! = \!\!\!\!\! & 
\frac{k_{3,3}^{5} (k_{2,5} k_{5,1} k_{1,9} k_{9,4} -
                            k_{4,9} k_{9,1} k_{1,5} k_{5,2})}
              {k_{5,5}^{6} k_{9,9}^{10} X} \nonumber \\
   & \!\!\!\!\! + \!\!\!\!\! & 
\frac{k_{3,3}^{5} (k_{2,5} k_{5,1} k_{1,9} k_{9,13} k_{13,10} 
                            k_{10,14} k_{14,4} -
                            \Omega_{R7B})}
              {k_{5,5}^{6} k_{9,9}^{10} k_{10,10}^{11} k_{13,13}^{14}
               k_{14,14}^{15} X} \nonumber \\
   & \!\!\!\!\! + \!\!\!\!\! & 
\frac{k_{3,3}^{5} (k_{2,5} k_{5,1} k_{1,9} k_{9,13} k_{13,4} -
                            \Omega_{R7C})}
              {k_{5,5}^{6} k_{9,9}^{10} k_{13,13}^{14} X}
       = 0.112 R1
\label{R7_1667}
\end{eqnarray}
where some reverse routes have been compressed to the \( \Omega \)-notation, as
for \( 1665 \)\,MHz.
For the equations in the set eq.(\ref{R2_1667}) - eq.(\ref{R4_1667}) where
there is more than one line, the lines are ordered by inverting strength. The
first line is termed the `A' route, then the `B' route. Only Route~7 has a
`C' route. In all these equations, inverting strengths have been given
relative to the expression in eq.(\ref{R1_1667}). The 
absolute value of this expression for \( R1 \) is
\( 6.89 \times 10^{-5} \)\,s\(^{-1}\). The seven routes represented by
eq.(\ref{R1_1667}) - eq.(\ref{R4_1667}) are shown on a schematic energy-level
diagram of OH in Fig.~\ref{f:fig1667}

\subsection{Summary}

The \( 1667 \)- and \( 1665 \)-MHz pumps share many features: the strongest
pair of routes are confined to the \( ^{2}\Pi_{3/2} \) stack of levels, the strongest
route does not involve levels higher than the \( ^{2}\Pi_{3/2}, J=5/2 \) rotational
state and the weaker routes make much more extensive use of the \( ^{2}\Pi_{1/2} \)
levels. The similarity is particularly striking for the strongest pumping
route in each line: for both lines, this consists (considering the forward
route only) of a radiative absorption from \( ^{2}\Pi_{3/2}, J=3/2 \) to \( ^{2}\Pi_{3/2}, J=5/2 \),
followed by a collisional de-excitation back to \( ^{2}\Pi_{3/2}, J=3/2 \). For
\( 1665 \)\,MHz, the route, in levels, is \( 1 \rightarrow 5 \), followed by
\( 5 \rightarrow 3 \). The analogous route at \( 1667 \)\,MHz is
\( 2 \rightarrow 6 \), followed by \( 6 \rightarrow 4 \). A similar analogy
can be drawn for the second-strongest route in both lines, where a 
collisional de-excitation is bracketed by a pair and a triplet of 
radiative lines. The first group operates between the lower halves of the
\( ^{2}\Pi_{3/2}, J=3/2 \) and \( ^{2}\Pi_{3/2}, J=5/2 \) lambda doublets, and the second group
links the upper halves. The swap from the lower to the upper halves is
allowed by the intervening collisional de-excitation.

\section{The Pumps at a Deeper Level}

So far, the present work has derived pumping routes for OH masers in terms of
all-process rate-coefficients. Therefore,
we know what routes are responsible for
most of the steady-state inversion, but some details of the underlying physics
are still missing. To proceed, we can expand the all-process rate coefficients
presented in eq.(\ref{R1_1665}) - eq.(\ref{R5_1665}) 
and eq.(\ref{R1_1667}) - eq.(\ref{R7_1667})
in terms of their radiative and collisional parts, so that 
the importance of routes can be
explained in terms of radiation fields and molecular parameters. Some of the
expressions are very complex, so I present here the analysis for the
most important route at \( 1665 \)\,MHz, and for its analogue at \(1667\)\,MHz,
where the analysis of the pumping schemes leads to a very simple physical
understanding. 

\subsection{1665\,MHz}

The strongest component of the \( 1665 \)\,MHz pump is given by the expression
in eq.(\ref{R1_1665}). The absolute value of the inversion
it generates requires multiplication of 
\( R1 \) by the
constants outside the main brackets in eq.(\ref{i1665}). Here, the only term
of interest is the antagonistic part of eq.(\ref{R1_1665}), so we group
the denominator with the constants and expand the expression,
\begin{equation}
y = k_{1,5} k_{5,3} - k_{3,5} k_{5,1}
\label{y0}
\end{equation}
As the \( 5 \rightarrow 3 \) transition is radiatively forbidden, expansion
of eq.(\ref{y0}) in terms of the radiation field and molecular parameters
yields
\begin{equation}
y = (B_{1,5} \bar{J}_{1,5} + C_{1,5}) C_{5,3} -
    C_{3,5} (A_{5,1} + \bar{J}_{1,5} B_{5,1} + C_{5,1})
\label{y1}
\end{equation}
where \( A_{x,y}\), (\(B_{x,y} \) ) are Einstein-A (B) coefficients and the
\( C_{x,y} \) are first-order collisional rate coefficients for population
transfer from level \( x \) to level \( y \). The values of the seven
important quantities appearing in eq.(\ref{y1}) are tabulated in
Table~\ref{bits1665}.
\begin{table}
\caption{Radiative and collisional rate coefficients for 1665\,MHz}
\begin{tabular}{@{}lr@{}}
\hline
Quantity                        &                 Value (Hz)         \\
\hline
\( A_{5,1} \)                   &  \(  1.24  \times 10^{-1} \)       \\
\( B_{5,1} \bar{J}_{1,5} \)     &  \(  2.16  \times 10^{-2} \)       \\
\( B_{1,5} \bar{J}_{1,5} \)     &  \(  3.60  \times 10^{-2} \)       \\
\( C_{5,1}               \)     &  \(  4.35  \times 10^{-5} \)       \\ 
\( C_{1,5}               \)     &  \(  1.31  \times 10^{-6} \)       \\
\( C_{5,3}               \)     &  \(  3.75  \times 10^{-4} \)       \\
\( C_{3,5}               \)     &  \(  1.13  \times 10^{-5} \)       \\
\hline
\end{tabular}
\label{bits1665}
\end{table}
It is easy to see from Table~\ref{bits1665} that there is a hierarchy of these
rate coefficients. The stimulated emission and absorption terms are about
five times smaller than the spontaneous emission rate, but of order \( 100 \)
times larger than the collisional rate coefficients. A sensible initial
simplification of eq.(\ref{y1}) is therefore to ignore terms that are
products of collisional coefficients, leaving
\begin{equation}
y \simeq B_{1,5} \bar{J}_{1,5} C_{5,3} -
    C_{3,5} (A_{5,1} + \bar{J}_{1,5} B_{5,1})
\label{y2}
\end{equation}
Another glance at Table~\ref{bits1665} shows that \( C_{3,5} \) is smaller
than \( C_{5,3} \) by a factor of about \( 30 \). Ignoring the stimulated
emission compared with the absorption reduces eq.(\ref{y2}) to
\begin{equation}
y \simeq B_{1,5} \bar{J}_{1,5} C_{5,3} - C_{3,5} A_{5,1}
\label{y3}
\end{equation}
The Einstein B-value and the upward collisional rate-coefficient can be
expressed in terms of the downward expressions, producing
\begin{equation}
y = (C_{5,3} g_{5}/g_{1}) [B_{5,1} \bar{J}_{1,5}  - A_{5,1} 
    e^{-\frac{h\nu}{kT_{K}}}]
\label{y4}
\end{equation}
where \( g_{1} \) and \( g_{5} \) are the statistical weights of levels \( 1 \)
and \( 5 \) respectively, \( \nu \) is the line-centre transition frequency
of the \( 5 \rightarrow 1 \) transition, and \( T_{k} \) is the kinetic
temperature. Expressing the Einstein
 B-coefficient in terms of the A-coefficient,
I obtain
\begin{equation}
y = \frac{A_{5,1} C_{5,3} g_{5}}{g_{1} (e^{\frac{h\nu}{kT_{K}}}-1)}
    \left[
      \frac{\bar{J}_{1,5}}{B_{\nu}(T_{K})} -1 + e^{-\frac{h\nu}{kT_{K}}}
    \right]
\label{y5}
\end{equation}
where \( B_{\nu}(T_{K}) \) is a Planck function at the local kinetic temperature
(for the chosen slab in the ALI model). The group of terms outside the
square braket in eq.(\ref{y5}) helps to set the overall speed of the pump, but
does not control its direction. For the moment, group these terms as \( K \),
and noting that the exponential in eq.(\ref{y5}) is vastly smaller than \( 1 \),
the final form for \( y \) is
\begin{equation}
y \simeq K [(\bar{J}_{1,5}/B_{\nu}(T_{K})) - 1]
\label{y6}
\end{equation}

Two things are responsible for the effectiveness of the pump in eq.(\ref{y6}).
The first is that the mean intensity of radiation in the \( 5 \rightarrow 1 \)
transition must be greater than a Planck function at the same frequency, based
on the kinetic temperature. The second is that the energy gap between the
\( ^{2}\Pi_{3/2}, J=3/2 \) and \( ^{2}\Pi_{3/2}, J=5/2 \) rotational levels is large compared
to \( kT_{K} \): this condition makes the upward rate coefficient in the
collision-only \( 5 \rightarrow 3 \) transition much weaker than its
downward companion.

The first of these requirements - that the mean intensity exceed the black-body
function at the kinetic temperature - can be explained in terms of three
physical processes, either singly or in combination. The first is the 
presence of a radiation field at a temperature higher than \( T_{K} \). The
ALI model has dust at a temperature \( T_{d} = 70 \)\,K, so continuum
radiation is a possible source of the necessary mean intensity. The second
possibility is that radiative transfer effects drive the mean intensity to
a level well above what would be expected for LTE at \(30\)\,K, the kinetic
temperature.
The third possibility is line
overlap with another FIR transition. However, in the model used here, the
\( 5 \rightarrow 1 \) and \( 6 \rightarrow 2 \) transitions are not members
of the overlapping groups, so the dust continuum and/or
NLTE radiation transfer must be responsible.

For the \( 1665 \)\,MHz pump, I compare the rate \( B_{5,1} \bar{J}_{5,1} \)
from Table~\ref{bits1665}
with three other relevant radiative rates. The first of these is the LTE
rate, \( B_{5,1} B_{\nu}(T_{K}) \), which is \( 2.28 \times 10^{-3} \)\,Hz
for \( T_{K} = 30\)\,K. The actual rate in the \( 5 \rightarrow 1 \) line
exceeds this by a factor of \( \sim 10 \). The second is the black-body
rate at the dust 
temperature, \( B_{5,1} B_{\nu}(T_{d}) = 2.70 \times 10^{-2}\)\,Hz - substantially larger than the actual rate. Thirdly, there is the rate
\( B_{5,1} S_{1,5} \) generated by the source function in the chosen slab,
which has the value \( 1.97 \times 10^{-2} \)\,Hz. The actual value exceeds
this slightly, so the actual radiative rate found in the transition must depend on
both a substantial optical
depth in the \( 5\rightarrow 1\) line, and the presence
of the dust continuum. This view of the inversion mechanism is consistent
with Fig.~\ref{inversionplot}, where there is little or no inversion for
models where \( T_{d} < T_{K} \).

It is, in fact, possible to give a reasonably detailed physical explanation of
the radiative part of the pump, that is to explain eq.(\ref{y6}) in terms of
the details of the radiative transfer. The ultimate source of the large
mean intensity in the \( 5 \rightarrow 1 \) line is the boundary condition
on the inner (high optical depth, and the more remote from the
observer) boundary of the model. This specifies that
the continuum becomes optically thick, and that the abundance of OH falls to
zero. As a result, the radiation field at the inner boundary is a Planck
function at the (constant) dust temperature, whilst the mean optical depth
in the line tends to a large, but finite, value. In the model analysed in
the present work, the mean line optical depth reaches \( \tau_{M} =
265 \). All the slabs which make up the numerical model (where
\( \tau < \tau_{M} \) ) will be referred to here as the line zone. The line
zone includes the slab analysed with {\sc tracer}. Although the line zone
also contains dust at the same temperature as the boundary, it is too optically
thin to contribute significant radiation to the pumping line. Therefore, it
is definitely the boundary which is responsible, and any simplified model
can assume that opacity is provided by the line only in the line zone, and
by the continuum only in the boundary. High line optical depths 
(\( > 150 \) ) in the analysed slab and those nearby, suggest that a diffusion
approximation may lead to a reasonable physical understanding of the line
zone close to the inner boundary.

If a radiation diffusion approximation is assumed (see Appendix~C), an 
analytical solution to the radiation transfer problem can be found. Within
the restrictions discussed in the Appendix, the mean intensity in the line
zone is given by,
\begin{equation}
\bar{J}(\tau ) = B_{\nu}(T_{K}) + (B_{\nu}(T_{d}) - B_{\nu}(T_{K}))
                 e^{-\sqrt{3\zeta}(\tau_{M} - \tau)}
\label{diffsoln}
\end{equation}
Equation \ref{diffsoln} has sensible limits: the mean intensity tends to a
Planck function at the dust temperature as the optical depth approaches the
inner boundary, whilst at small optical depth, the limit is a Planck function
at the kinetic temperature. Of course eq.(\ref{diffsoln}) has no semblance
of validity for small values of \( \tau \). Re-arranging eq.(\ref{diffsoln})
to look like eq.(\ref{y6}) yields,
\begin{equation}
\frac{\bar{J}(\tau )}{B_{\nu}(T_{K})} - 1 =
\left(
\frac{B_{\nu}(T_{d})}{B_{\nu}(T_{K})} - 1
\right) e^{-\sqrt{3\zeta}(\tau_{M} - \tau)}
\label{last1}
\end{equation}
so in the model used in the present work, the boundary value of the bracket
( \( 10.87 \) ) has decayed to \( 8.496 \) over an optical depth shift of
\( 32.4 \). This shows that the crucial parameter is \( \zeta \). If it were
not present in eq.(\ref{last1}), the mean intensity would decay to a value
dictated by the kinetic temperature well within the first slab of the line
zone, and no pumping radiation would be transported far from the inner
boundary. The full definition of \( \zeta \), the scattering parameter, is
given in Appendix~C for a two-level system. However, in the case of the
\( 5 \rightarrow 1 \) line in the present work, with parameters given in
Table~\ref{bits1665}, it is well approximated by 
\( \zeta \sim C_{5,1}/A_{5,1} = 3.51\times 10^{-4} \). This results in an
effective optical depth scale thinner than that based on the line-profile
mean by a factor of \( \sqrt{3 \zeta } = 0.032 \). This scale does not match
the decay in \( \bar{J} \) found in the model, but this is not surprising
given that the diffusion approximation is not strictly valid, and relies
on the assumption of a two-level system. The important point is that the
effective optical depth scale is vastly thinner than the line scale because
of a small value of \( \zeta \). The small value of \( \zeta \) in turn 
depends upon the fact that the Einstein A-value in the \( 5 \rightarrow 1 \)
line is vastly larger than the downward collisional rate coefficient. Any
radiation absorbed in this line has a high probability of being re-radiated,
in the same line, rather than being collisionally de-excited, which would
lead to thermalization of the energy. A large fraction of the absorption
therefore behaves effectively as scattering, so that the effects of the
boundary continuum on pumping can be felt in the slab studied with {\sc tracer},
and indeed substantially closer to the observer.

It is perhaps worth noting that photons travelling directly from the boundary
do not actually penetrate very far into the model. This radiation also does not decay
exponentially as \( \tau_{M} - \tau \), because the wings of the Gaussian line
profile are always substantially optically thinner than the line mean. Very
roughly, the actual inner boundary radiation penetrates as
\( 1/(\tau_{M} - \tau ) \), for optical
depth shifts \( > 1 \),
so it would have only a marginal effect on the
slab studied in the present work, when compared to the `effectively
scattered' radiation discussed above. 

An interesting rider to this investigation of the leading \( 1665\)-MHz pump
is to consider the symmetry between 
the routes linking levels \( 3 \) and \( 1 \)
via level \( 5 \), as discussed above, and those linking levels \( 3 \) and
\( 1 \) via level \( 7 \). If one assumes similar Einstein coefficients and
collisional rate coefficients, one would conclude that a radiative excitation
from level \( 3 \) to level \( 7 \), followed by a collisional decay from
\( 7 \) to level \( 1 \), would provide an anti-inverting mirror to the
similar transitions via level \( 5 \), negating the inverting effect of
the latter. The reason that this does not happen in the model discussed
here is that the downward rate-coefficient, 
\( C_{7,1} = 5.00 \times 10^{-5}\)\,Hz, is considerably smaller (by a factor
of \( 0.133 \) )
than \( C_{5,3} \) (see Table~\ref{bits1665}). Importantly, \( C_{5,3} \) is
the largest term in the {\sc tracer} expansion of \( k_{5,3}^{6} \), which
follows from eq.(\ref{R1_1665}), and the fact that the \( 5 \rightarrow 3 \)
transition is radiatively forbidden. Therefore, direct collisional transfer
in this route exceeds all indirect methods of transfer between levels \( 5 \)
and \( 3 \), and the large difference between \( C_{5,3} \) and \( C_{3,5} \)
is significant in creating inversion. By contrast, \( C_{7,1} \) is only the
third most important route in the expansion of \( k_{7,1}^{8} \). It is
exceeded in importance by radiative routes transfering population from level
\( 7 \) to level \( 1 \) via levels in the \( ^{2}\Pi_{1/2} \) stack.
Therefore, compared with \( 5 \rightarrow 3 \), direct collisional transfer
between level \( 7 \) and level \( 1 \) plays a much less prominent role, and
the difference between \( C_{1,7} \) and \( C_{7,1} \), though large, is
unimportant as an anti-inverting mechanism.

At a still deeper level, it is possible to explain why the collisional
rate coefficient \( C_{7,1} \) is so much smaller than \( C_{5,3} \): both
coefficients are constructed from contributions due to collisions with 
ortho- and para-H\(_{2}\). The rate coefficients from collisions of OH with
ortho-H\(_{2}\) are similar for both transitions, with the coefficient for
\( 7 \rightarrow 1 \) stronger by a factor of \( 1.08 \). 
However the rate coefficients for the two transitions derived from collisions
with para-H\(_{2}\) are very different. This
`parity propensity' of the OH plus para-H\(_{2}\) collision cross sections
was noted by Offer et al. \shortcite{off94}. In fact, the ratio of the
rate coefficients for the \( 5 \rightarrow 3 \) and \( 7 \rightarrow 1 \)
transitions, due to collisions with para-H\(_{2}\), is \( 8.37 \) at the
\( 30 \)\,K kinetic temperature of the model. The strong parity propensity of
the collisions with para-H\(_{2}\) is apparent at this temperature because
the equilibrium abundance of the ortho-H\(_{2}\) is small (about \( 2.5 \)
per cent). At higher temperatures, the abundance of ortho-H\(_{2}\) would
be expected to rise, and the efficiency of the pump to fall, as the
ortho- to para-H\(_{2}\) abundance ratio tends towards its high-temperature
value of \( 3 \). A glance at Fig.~\ref{inversionplot} suggests
that this is indeed the case, with the pumping mechanism becoming 
ineffective below a kinetic temperature of \( 100 \)\,K, where the 
ratio of the contributions of the ortho-
and para-hydrogen species to the overall rate coefficient is \( 1.5 \).

\subsection{1667\,MHz}

The strongest pump route at \( 1667 \)\,MHz can be analysed in a very similar
way to the \( 1665 \)\,MHz route 
discussed above. The inversion expression to be
expanded is 
\begin{equation}
y' = k_{2,6} k_{6,4} - k_{4,6} k_{6,2}
\label{yp0}
\end{equation}
and the version of eq.(\ref{yp0}) fully expanded in terms of molecular
parameters and the radiation field in transition \( 2 \rightarrow 6 \) is
\begin{equation}
y' = (B_{2,6} \bar{J}_{2,5} + C_{2,6}) C_{6,4} -
    C_{4,6} (A_{6,2} + \bar{J}_{2,6} B_{6,2} + C_{6,2})
\label{yp1}
\end{equation}
Values of the various terms appearing in eq.(\ref{yp1}) appear in
Table~\ref{bits1667}.
\begin{table}
\caption{Radiative and collisional rate coefficients for 1667\,MHz}
\begin{tabular}{@{}lr@{}}
\hline
Quantity                        &                 Value (Hz)         \\
\hline
\( A_{6,2} \)                   &   \( 1.38  \times 10^{-1} \)       \\
\( B_{6,2} \bar{J}_{2,6} \)     &   \( 2.31  \times 10^{-2} \)       \\
\( B_{2,6} \bar{J}_{2,6} \)     &   \( 3.24  \times 10^{-2} \)       \\
\( C_{6,2} \)                   &   \( 5.16  \times 10^{-5} \)       \\ 
\( C_{2,6} \)                   &   \( 1.30  \times 10^{-6} \)       \\
\( C_{6,4} \)                   &   \( 4.74  \times 10^{-4} \)       \\
\( C_{4,6} \)                   &   \( 1.20  \times 10^{-5} \)       \\
\hline
\end{tabular}
\label{bits1667}
\end{table}
Following through the same steps as for \(1665\)\,MHz, the final expression for
\(y'\) is
\begin{equation}
y' \simeq K' [(\bar{J}_{2,6}/B_{\nu}(T_{K})) - 1]
\label{yp2}
\end{equation}
It can be shown that the mean intensity in the \( 2 \rightarrow 6 \) line 
which pumps \( 1667 \)\,MHz is actually smaller than the corresponding
mean intensity in the \( 1 \rightarrow 5 \) transition. On this basis \( y' \)
in eq.(\ref{yp2}) would be expected to be smaller than \( y \) for the
case of identical constants \( K = K' \). The reason that the \( 1667 \)\,MHz
inversion is larger, per sublevel, by a factor of \( 1.188 \) therefore
requires that either \( K > K' \), or that the inversion ratio is explained
by differences in the external constants appearing
in eq.(\ref{i1665}) and eq.(\ref{i1667}). If we take the inversion per
sublevel, then the ratio of the external constants is
\begin{equation}
R_{ext} = \frac{ \Delta \rho_{42}}{ \Delta \rho_{31}} =
        \frac{3(k_{1,1}^{5} k_{3,3}^{5} - k_{1,3}^{5} k_{3,1}^{5})}
             {5 k_{2,2}^{4} k_{4,4}^{5}}
\label{external}
\end{equation}
When evaluated, \( R_{ext} = 1.093 \). Therefore, part of the reason for the
larger inversion at \( 1665 \)\,MHz lies in the ratio of the constants,
\( R_{c} = K'/K \). Ignoring the exponential terms, the ratio of which is
\( 1.0 \) to better than one part in \( 10^{4} \), \( R_{c} \) is given by
\begin{equation}
R_{c} = \frac{A_{6,2} C_{6,4} g_{6} g_{1}}
             {A_{5,1} C_{5,3} g_{2} g_{5}}
\label{internal}
\end{equation}
which evaluates to \( R_{c} = 1.1794 \). One reason why the inversion at
\( 1667 \)\,MHz is larger than the corresponding inversion at \( 1665\)\,MHz
is therefore that both transitions in its strongest pumping routes are
faster than their analogues in the \( 1665 \)\,MHz pump. For comparison of
the Einstein A-values and collisional rate coefficients, see
Table~\ref{bits1667} and Table~\ref{bits1665}. The radiative part of the
pump for \( 1667 \)\,MHz depends on the efficient transfer of radiation
from the inner boundary of the model by effective scattering of radiation in
the \( 6 \rightarrow 2 \) line in a very similar manner to the detailed 
discussion for \( 1665 \)\,MHz, given in Section~6.1.

The similarity of the \( 1667 \)- and \( 1665 \)-MHz pumps also extends to
the symmetry breaking between the pumping route (level \( 2 \) to level \( 4 \)
via level \( 6 \)) investigated above, and its anti-inverting mirror, taking
population from level \( 4 \) to level \( 2 \) via level \( 8 \). Just as in
the case of the \( 1665 \)-MHz pump, the inverting route has the direct
collisional transfer as the leading term in the {\sc tracer} expansion of its
downward transition ( \( 6 \rightarrow 4 \) in this case), but the direct
transfer is relatively unimportant in the downward transition
( \( 8 \rightarrow 2 \) ) of the anti-inverting route. The weakness of the
direct \( 8 \rightarrow 2 \) collisional transfer is based on the size of
the collisional rate coefficient (\( C_{8,2}/C_{6,4} = 0.18 \)), which in
turn depends on the overwhelming contribution of the para-H\(_{2}\) species,
with its large parity propensity,
to collisions at low temperatures.

\section{Efficiencies}

I discuss here the efficiency of overall schemes and the
efficiencies of important individual routes. I define the efficiency of a
scheme or route as the ratio of the rate at which it produces inversion
divided by the total population flow rate through the same scheme or route.
For example, considering the overall \( 1665 \)\,MHz pumping scheme, there
are the direct and indirect terms from eq.(\ref{i1665}). For the direct part
of the pump, the efficiency is
\begin{equation}
\epsilon_{d} = \frac{k_{1,3}^{4}-k_{3,1}^{4}}{k_{1,3}^{4}+k_{3,1}^{4}}
\label{eff0}
\end{equation}
which evaluates to \( 4.1 \) per cent. Similarly, the
indirect part of the pump is \( 3.34 \) per cent efficient. Weighting these
two efficiencies by the contribution each makes to the inversion, the
overall efficiency of the \( 1665 \)\,MHz pump is \( 3.75 \) per cent. In the
case of the \( 1667 \)\,MHz pump, obtaining the overall efficiency is a
little more laborious, but still straightforward, and the mean of all the
routes in eq.(\ref{i1667}), weighted by contribution to the inversion gives
an efficiency of \( 4.60\) per cent. The third term in the brackets of
eq.(\ref{i1667}) is notable in having the largest individual efficiency, of
\( 5.13\) per cent. Overall, the \( 1667\)\,MHz pump is the more efficient,
which is unsurprising, since the inversion in this line is also the larger.

\subsection{individual routes}

Individual routes within the overall scheme can have substantially higher
efficiencies than those for the scheme as a whole. I define such internal
efficiencies as those calculated for a route when the denominator is
confined to consideration of that route only. For example, the strongest
pumping route at \( 1665 \)\,MHz has the internal efficiency,
\begin{equation}
\epsilon_{R1} = \frac{k_{1,5} k_{5,3} - k_{3,5} k_{5,1}}
                     {k_{1,5} k_{5,3} + k_{3,5} k_{5,1}}
\label{eff1}
\end{equation}
which has the numerical value of \( 78.2 \) per cent. The analagous route in
the \( 1667\)\,MHz system, operating via levels \( 2\), \( 6 \) and \( 4 \)
is internally \( 77.6 \) per cent efficient. These figures drop to 
\( \sim 2.5 \) per cent when the routes are expressed as part of the
expansion of eq.(\ref{eff0}), or its \( 1667 \)\,MHz analogue. Routes with
many links are not noticeably less efficient. For example, Route~2B in
the \( 1665 \)\,MHz scheme, which climbs to level \( 20 \) in the
\( ^{2}\Pi_{3/2}, J=7/2 \) rotational level, has the internal efficiency
\begin{equation}
\epsilon_{R2B} = 
\frac{k_{1,5} k_{5,2} k_{2,6} k_{6,4} k_{4,8} k_{8,20} k_{20,7} k_{7,3}-
              \Omega_{2B}}
     {k_{1,5} k_{5,2} k_{2,6} k_{6,4} k_{4,8} k_{8,20} k_{20,7} k_{7,3}+
              \Omega_{2B}}
\label{eff2}
\end{equation}
with the numerical value of \( 77.0 \) per cent.

\section{Discussion}

The results reported here and in Gray et al. \shortcite{ghl05} show that it
is possible to trace pumping schemes in OH in considerable detail in at least
two of the common OH-maser environments (Galactic star-forming regions and
evolved-star envelopes). The pumping schemes revealed so far are actually
very complex, but a large proportion of their inverting strength can be
explained in terms of a few dominant routes. An obvious extension of this
work is to study the additional ground-state OH maser environments: 
OH megamasers and those supernova remnants which support strong masing at
\( 1720 \)\,MHz. Additional work needs to be done to confirm the generality
of the schemes analysed in the present work, both in terms of the spatial
variation within one model, and variation between models with different
input parameters. To complete such an analysis of many schemes in a reasonable
time requires considerably improved automation of the population tracing
process. The explanation of the radiative part of the pump in terms of the
transport of radiation from the optically thick boundary to a significant
part of the model (many slabs) by means of a modifed optical depth scale does
at least suggest that the same pumping mechanism operates within a large
fraction of the current geometrical model.

The importance of parity propensity in collisions between OH and para-H\(_{2}\)
in the pumping scheme for both OH main lines means that the ortho- to
para-H\(_{2}\) ratio in the gas of the maser zone must be well biased in
favour of the para-species for the schemes investigated here to operate
efficiently. Given that H\(_{2}\) forms in the ratio of 3:1 in favour of
the ortho-form, as dictated by the nuclear spin statistics, maser action via
the pumping schemes discussed here would require time for the species ratio
to move substantially towards the thermodynamic ratio via a set of proton
exchange reactions \cite{fpw06}. In some circumstances the timescale for
these reactions could equate to the `switch-on' time for main-line OH masers.

Labelling of energy levels is of course arbitrary, so expressions similar to
eq.(\ref{i1665}) and eq.(\ref{i1667}) can be formed for excited-state
inversions without increased complexity, providing that, say, the
\( ^{2}\Pi_{3/2}, J = 5/2 \) levels were labelled \( 1-4 \) instead of those in the
\( ^{2}\Pi_{3/2}, J = 3/2 \) ground state. It would then be possible to trace the
common \( 6030 \) and \( 6035 \)\,MHz inversions; alternative re-labelling
would allow {\sc tracer } to be applied to any of the other OH rotational
levels that support maser action. A likely problem with the extension of the
method to excited states is one of stability: the elimination process 
following the naive method relies on the last levels eliminated having large
populations as a form of numerical pivoting. There may, however, be ways of
avoiding these stability problems by attaching the operation log to the
basic numerical method, provided that the interpretation in terms of rate
coefficients can be preserved.

It may also be possible to extend this inversion-tracing method to
molecules other than OH. In many interesting cases, such as the \(22\)\,GHz
maser line of water, the maser levels lie well above the ground state, so
any attempts at analysis would suffer from the stability problems already
discussed in connexion with the excited rotational states of OH. However, 
there may be an even more fundamental barrier to understanding inversions
in some cases. In the case of methanol \cite{sobdeg94} the authors show
that, for the \(2_{0}-3_{-1}\) Class II maser transition
of the E-species, \(65\) per cent of the maser flux results from
pumping cycles with a number of links which exceeds \(10\). In these cases
at least, it is almost certainly not possible to write down a manageable
expression for the inversion: either the number of required terms would be
too great, or the expressions too complicated, or both. If the {\sc tracer}
method is to be extended to another molecule, the first one to try is
probably SiO: it has a relatively simple level structure, and a maser (though
not one of the strongest or most common) has been observed between the
lowest two rovibrational levels (\( v=0, J=1-0\)). In addition, population in
each vibrational level tends to be concentrated towards the lower rotational
levels, which is likely to aid numerical stability.

Another application which is likely to prove fruitful is to extend the 
{\sc tracer} method to saturating conditions. The semi-classical analysis
by Field \& Gray \shortcite{fg88} automatically breaks the kinetic scheme
into coefficients which multiply the maser radiation and those which do
not. The former affect the inversion as the maser saturates, whilst the latter
generate the inversion, like those studied in the present work. Both sets
of coefficients in models (for example Gray, Field \& Doel \shortcite{gfd92})
are generated in a very similar manner, and, for the ground-state OH masers,
would introduce no extra stability problems. The coefficients that multiply
the maser intensity could be studied as a function of propagation distance
along the maser, showing how the increasingly rapid transfer of population
across the maser levels introduces newly important transfer routes.

\section{Conclusions}

I conclude that it is possible to trace the population transfer routes
which cause strong inversions in the \(18\)\,cm OH main-line masers under
conditions typical of star-forming regions. The number of routes which need
to be traced in order to recover the great majority of the inversion is
modest - a conclusion which does not necessarily hold for the kinetic
schemes of other molecules, or even for OH in other environments.

The pumping schemes derived for the \( 1665 \)- and \( 1667\)-MHz lines show
strong similarity, and are biased quite strongly to the \(^{2}\Pi_{3/2}\) stack of
rotational levels, unlike the case of \( 1612 \)\,MHz OH masers in
late-type stellar atmospheres. In both lines, the strongest route contains
just two transitions, comprising, for the forward route, a radiative
absorption, lifting population to the \( ^{2}\Pi_{3/2}, J=5/2 \) rotational state,
followed by collisional de-excitation to the upper part of the ground-state
lambda doublet. This inclusion of a radiatively forbidden link to switch
from the lower to the upper half of the lambda-doublet system is also
employed by the next most important pumping route in both main lines.
Anti-inverting `mirror' transitions are not effective because of the
parity propensity found in collisions between OH and para-H\(_{2}\).

The effectiveness of these routes in the \(^{2}\Pi_{3/2}\) stack is based firstly upon the
fact that the energy gap between the \( ^{2}\Pi_{3/2}, J=3/2 \) and \( ^{2}\Pi_{3/2}, J=5/2 \)
rotational states is large compared to the energy equivalent of the kinetic
temperature, which favours collisional de-excitation over excitation, and
secondly on the presence of a dust radiation field which has a characteristic 
temperature which is locally hotter than the kinetic temperature, a result
of efficient transfer of radiation from the optically thick boundary by
`effective scattering'.
  
\subsection*{ACKNOWLEDGMENTS}

MDG acknowledges PPARC for financial support under the UMIST astrophysics
2002-2006 rolling grant, number PPA/G/O/2001/00483, and thanks the referee for
helpful suggestions regarding the improvement of the later parts of the paper.

\appendix

\section{Diagonal Coefficients}

The aim of this section is to prove that a diagonal coefficient,
\( k_{j,j}^{p} \), at 
elimination stage \( p \) is equal to the sum of the all-process 
rate-coefficients from level \( j \) to all the un-eliminated levels, or
\begin{equation}
k_{j,j}^{p} = \sum_{m=1; m \neq j}^{p-1} k_{j,m}^{p}
\label{sumk}
\end{equation}
This is true by definition for the original matrix (\( p=N+1\)), but is not
obviously so for any smaller value of \( p \).

The state of the set of kinetic
master equations at elimination stage \( p \) looks like:
\begin{equation}
   \begin{array}{ccccccr}
k_{p-1,p-1}^{p} \rho_{p-1} \!\!\!\! &
  \!\!\!\! ... \!\!\!\! & \!\!\! ... & \!\!\!\!\!\! ... \!\!\!\!
& \!\!\! -k_{i,p-1}^{p} \rho_{i} \!\!\!\! & 
\!\!\!\!\!\!\!\!\! ... \!\!\!\!\! & \!\!\!\!\! -k_{1,p-1}^{p} \rho_{1} \!=\! 0\\
-k_{p-1,p-2}^{p} \rho_{p-1} \!\!\!\!&
  \!\!\!\! ...  \!\!\!\!  & \!\!\!... & \!\!\!\!\!\! ... \!\!\!\!
& \!\!\! -k_{i,p-2}^{p} \rho_{i} \!\!\!\! & 
\!\!\!\!\!\!\!\!\! ... \!\!\!\!\! & \!\!\!\!\! -k_{1,p-2}^{p} \rho_{1} \!=\! 0\\
\vdots                       \!\!\!\!  &
  \!\!\!\! ... \!\!\!\!   & \!\!\!... & \!\!\!\!\!\! ... \!\!\!\!
& \!\!\! ...                    \!\!\!\! & 
\!\!\!\!\!\!\!\!\! ... \!\!\!\!\! & \!\!\!\!\! \vdots                 = \!0\! \\
- k_{p-1,j}^{p} \rho_{p-1} \!\!\!\! & 
 \!\!\!\! ...  \!\!\!\!   & \!\!\!
 + k_{j,j}^{p} \rho_{j} & \!\!\!\!\!\! ... \!\!\!\!
& \!\!\! -k_{i,j}^{p} \rho_{i}  \!\!\!\! & 
 \!\!\!\!\!\!\!\!\! ... \!\!\!\!\! & \!\!\!\!\! -k_{1,j}^{p} \rho_{1} = \!0\! \\
\vdots                       \!\!\!\! & 
\!\!\!\! ... \!\!\!\! & \!\!\! ... & \!\!\!\!\!\! ... \!\!\!\!
& \!\!\! ...                   \!\!\!\!  & 
\!\!\!\!\!\!\!\!\! ... \!\!\!\!\! & \!\!\!\!\! \vdots                 = \!0\! \\
-k_{p-1,2}^{p} \rho_{p-1} \!\!\!\! & 
 \!\!\!\! ...  \!\!\!\!   & \!\!\! ... & \!\!\!\!\!\! ... \!\!\!\!
& \!\!\! ...                    \!\!\!\! & 
\!\!\!\!\!\!\!\!\! +k_{2,2}^{p} \rho_{2} \!\!\!\!\! & \!\!\!\!\! -k_{1,2}^{p} 
\rho_{1} = \!0\!\\ 
 k_{p-1,1}^{p*} \rho_{p-1} \!\!\!\!&
 \!\!\!\! ... \!\!\!\!    & \!\!\! ... & \!\!\!\!\!\! ... \!\!\!\!
& \!\!\! ...                   \!\!\!\!  & 
\!\!\!\!\!\!\!\!\! +k_{2,1}^{p*} \rho_{2} \!\!\!\!\! & 
\!\!\!\!\! +k_{1,1}^{p*}\rho_{1} \!=\!
{\cal N} \\
   \end{array}
\label{matrix}
\end{equation}
where \( \rho_{x} \) is the population of level \( x \) and the 
conservation equation coefficients, written with a star in the superscript,
appear in the final equation. An additional equation is now eliminated by
the naive method: the topmost equation in the set eq.(\ref{matrix}) is
used to eliminate \( \rho_{p-1} \), such that
\begin{equation}
\rho_{p-1} = (\sum_{m=1}^{p-2} k_{m,p-1}^{p} \rho_{m} )/k_{p-1,p-1}^{p}
\label{rhop-1}
\end{equation}
and this population is eliminated from all the others in favour of the
expression in eq.(\ref{rhop-1}). Concentrating on the arbitrary equation
\( j \) such that \( j \leq p-2 \), the diagonal coefficient in this equation
is modified to the form
\begin{equation}
k_{j,j}^{p-1} = k_{j,j}^{p} - \frac{k_{p-1,j}^{p} k_{j,p-1}^{p}}
                                   {k_{p-1,p-1}^{p}}
\label{modiag}
\end{equation}
I now assume that the desired result is true at elimination stage \( p \),
and prove by induction that it is true at all 
further stages with \( p > 2 \), noting that
at \( p = 3 \), only a trivial \( 2 \times 2 \) matrix remains. This 
assumption allows the development of eq.(\ref{modiag}) to
\begin{eqnarray}
k_{j,j}^{p-1} &=& k_{j,1}^{p} + ... + k_{j,j-1}^{p} + k_{j,j+1}^{p} + ... +
                k_{j,p-1}^{p} \nonumber \\
              & - &k_{p-1,j}^{p} k_{j,p-1}^{p}/k_{p-1,p-1}^{p}
\label{modmodiag}
\end{eqnarray}
I now add zero to each term on the right-hand side of eq.(\ref{modmodiag}), but
in a form which allows development of the off-diagonal coefficients.
Equation \ref{modmodiag} becomes,
\begin{eqnarray}
k_{j,j}^{p-1} &=& k_{j,1}^{p} + \frac{k_{j,p-1}^{p} k_{p-1,1}^{p}}
                                     {k_{p-1,p-1}^{p}}
                               - \frac{k_{j,p-1}^{p} k_{p-1,1}^{p}}
                                     {k_{p-1,p-1}^{p}} 
\nonumber \\
          + &...& +  k_{j,j-1}^{p} + \frac{k_{j,p-1}^{p} k_{p-1,j-1}^{p}}
                                              {k_{p-1,p-1}^{p}}
                                        - \frac{k_{j,p-1}^{p} k_{p-1,j-1}^{p}}
                                              {k_{p-1,p-1}^{p}}
\nonumber \\
            & + & k_{j,j+1}^{p} + \frac{k_{j,p-1}^{p} k_{p-1,j+1}^{p}}
                                              {k_{p-1,p-1}^{p}}
                                  - \frac{k_{j,p-1}^{p} k_{p-1,j+1}^{p}}
                                              {k_{p-1,p-1}^{p}}
\nonumber \\
          + &...& + k_{j,p-2}^{p} + \frac{k_{j,p-1}^{p} k_{p-1,p-2}^{p}}
                                              {k_{p-1,p-1}^{p}}
                                  - \frac{k_{j,p-1}^{p} k_{p-1,p-2}^{p}}
                                              {k_{p-1,p-1}^{p}}
\nonumber \\
           & + & k_{j,p-1}^{p} - \frac{k_{j,p-1}^{p} k_{p-1,j}^{p}}
                                              {k_{p-1,p-1}^{p}}
\label{bigdiag}
\end{eqnarray}
and the off-diagonal coefficients are now of a form which can be upgraded to
the next elimination stage via eq.(\ref{coef}). The coefficient at level
\( p \) combines with the positive fraction on each line (except the last)
to form a coefficient at stage \( p - 1 \), leaving
\begin{eqnarray}
k_{j,j}^{p-1} &\!=\!& 
k_{j,1}^{p-1}+ ... + k_{j,j-1}^{p-1} +k_{j,j+1}^{p-1} + ... +
                  k_{j,p-2}^{p-1} + k_{j,p-1}^{p} 
\nonumber \\
              &\!-\!& 
\frac{k_{j,p-1}^{p}}{k_{p-1,p-1}^{p}}
\left(
 k_{p-1,1}^{p} + ... + k_{p-1,j-1}^{p}  \right. \nonumber \\
 & \!+\! & 
\left. k_{p-1,j+1}^{p} + ... + k_{p-1,p-2}^{p}
+ k_{p-1,j}^{p}
\right)
\label{p-1diag}
\end{eqnarray}
where a common factor of \(k_{j,p-1}^{p}/k_{p-1,p-1}^{p}\) has been extracted
from the negative terms on the right-hand side of eq.(\ref{bigdiag}). I now
note that the bracket in eq.(\ref{p-1diag}) is the sum of all the rate
coefficients taking population out of level \( p-1 \), and is therefore equal
to the diagonal coefficient \( k_{p-1,p-1}^{p} \), given the assumption about
such coefficients at elimination stage \( p \). The bracket therefore cancels
with this coefficient in the denominator, leaving two coefficients of the
form \( k_{j,p-1}^{p} \) with opposite signs. Cancellation of these leaves
\begin{equation}
k_{j,j}^{p-1} = \sum_{m=1;m \neq j}^{p-2} k_{j,m}^{p-1}
\label{qed}
\end{equation}
which is of the form of eq.(\ref{sumk}), but with \( p \) 
replaced by \( p - 1\). Therefore, if the initial assumption is true for any
given stage of elimination, it remains true for the next. As it is true by
definition for the original set of equations, where \( p = N+1 \), then the
assumption is true for all subsequent eliminations at least as far as
\( p = 3 \).

\section{Denominator}

This section shows that the external denominator (\( D \) in 
eq.(\ref{i1665}) and eq.(\ref{i1667})) is positive definite. In terms of
rate coefficients evaluated at elimination stage \( p=4 \), the
denominator is given by
\begin{eqnarray}
D & = & k_{3,3}^{4} (k_{1,1}^{4*} k_{2,2}^{4} + k_{1,2}^{4} k_{2,1}^{4*}) +
        k_{3,1}^{4*}(k_{1,3}^{4}  k_{2,2}^{4} + k_{1,2}^{4} k_{2,3}^{4} ) 
    \nonumber \\
  & + &
        k_{3,2}^{4} (k_{1,3}^{4} k_{2,1}^{4*} - k_{1,1}^{4*} k_{2,3}^{4})
\label{Dapp0}
\end{eqnarray}
where coefficients marked with an asterisk in the superscript 
(starred) come from the
conservation equation,
\begin{equation}
\sum_{i=1}^{p-1} k_{i,1}^{p*} \rho_{i} = {\cal N}
\end{equation}
and cannot be interpreted in the same manner as the other rate-coefficients
via eq.(\ref{coef}). At \( p = N+1 \), these starred coefficients are all
equal to \( 1.0 \) and subsequent actions can only add combinations of
positive-definite rate-coefficients to them. They are therefore positive
definite at any value of \( p \). Given this result, and the proof in 
Appendix~A, the middle term in eq.(\ref{Dapp0}) is positive definite, and
will be called \( B \). Using the result of Appendix~A to expand
\( k_{3,3}^{4} = k_{3,1}^{4} + k_{3,2}^{4} \), eq.(\ref{Dapp0}) becomes
\begin{eqnarray}
D& \!\!\!\!\! =  \!\!\!\!\! &
k_{3,1}^{4} k_{1,1}^{4*} k_{2,2}^{4}+k_{3,1}^{4} k_{1,2}^{4} k_{2,1}^{4*}
  + k_{3,2}^{4} k_{1,1}^{4*} k_{2,2}^{4} + k_{3,2}^{4} k_{1,2}^{4} k_{2,1}^{4*}
\nonumber \\
 & \!\!\!\!\! + \!\!\!\!\! &B 
  + k_{3,2}^{4} k_{1,3}^{4} k_{2,1}^{4*} - k_{3,2}^{4} k_{1,1}^{4*} k_{2,3}^{4}
\label{Dapp1}
\end{eqnarray}
following the expansion of the first and third brackets of eq.(\ref{Dapp0}).
The first, second and fourth terms
of eq.(\ref{Dapp1}), and the first term following \( B \),
are all positive definite, and may be combined as \( A \). This leaves,
\begin{equation}
D = A + k_{3,2}^{4} k_{1,1}^{4*} k_{2,2}^{4} + B
      - k_{3,2}^{4} k_{1,1}^{4*} k_{2,3}^{4}
\label{Dapp2}
\end{equation}
but when \( k_{2,2}^{4} \) is expanded in accordance with Appendix~A, it
contains \( k_{2,3}^{4} \), so the remaining negative term in 
eq.(\ref{Dapp2}) is cancelled exactly, and \( D \) is positive definite.

\section{Radiation Diffusion}

In a zone where radiation transfer is diffusive, application of the
Eddington approximation (see for example Rybicki \& Lightman \shortcite{rl79})
leads to a radiation diffusion equation for the mean intensity of the form,
\begin{equation}
\frac{d^{2}\bar{J}}{d\tau^{2}} = 3 \zeta ( \bar{J}(\tau) - B_{\nu}(\tau) ),
\label{radodiffn}
\end{equation}
where \( B_{\nu}(\tau) \) is a the Planck function 
at the kinetic temperature, and the source function has
been elimated in favour of the mean intensity, using the standard expression
from molecular kinetics,
\begin{equation}
\bar{S}(\tau) =\zeta (\tau ) B_{\nu}(\tau) + (1 - \zeta (\tau) ) \bar{J}(\tau ).
\label{kintex}
\end{equation}
I note that eq.(\ref{kintex}) implicity assumes a two-level model: in the 
many-level case, the multiplier of \( \bar{J} \) cannot be represented as
\( 1 - \zeta \), and both this multiplier and \( \zeta \) depend on the
mean intensities and molecular parameters of all other radiative and 
collisionally allowed transitions. In the two-level case, the parameter
\( \zeta \) is independent of optical depth if the kinetic
temperature is constant, and is given by
\begin{equation}
\zeta = \frac{g_{u} C_{u,l} - g_{l} C_{l,u}}
             {g_{u} C_{u,l} - g_{l} C_{l,u} + g_{u} A_{u,l}}
\label{zetadef}
\end{equation}
where \( u \) is the upper and \( l \), the lower level of the transition.
As elsewhere in the paper, \( C_{u,l} \) ( \( C_{l,u} \) ) is the downward
(upward) collisional rate coefficient, \( A_{u,l} \) is the Einstein 
A-coefficient and \( g_{u} \), \( g_{l} \) are statisical weights. 


\begin{thebibliography}{}
\bibitem[\protect\citename{Andresen, Hausler \& L\"{u}lf }1984]{ahl84}
Andresen P., Hausler D., L\"{u}lf H.W., 1984, A\&A, 138, 17
\bibitem[\protect\citename{Bujarrabal et al. }1980]{buj80}
Bujarrabal V., Guibert J., Nguyen-Q-Rieu, Omont A., 1980, A\&A, 84, 311
\bibitem[\protect\citename{Cesaroni \& Walmsley }1991]{cw91}
Cesaroni R., Walmsley C.M., 1991, A\&A, 241, 537
\bibitem[\protect\citename{Collison \& Nedoluha }1993]{cn93}
Collison A.J., Nedoluha G.E., 1993, ApJ, 413, 735
\bibitem[\protect\citename{Collison \& Nedoluha }1994]{cn94}
Collison A.J., Nedoluha G.E., 1994, ApJ, 422, 193
\bibitem[\protect\citename{Destombes et al. }1977]{Destombes}
Destombes J.L., Marli\`{e}re C., Baudry A., Brillet J., 1977, A\&A, 60, 55
\bibitem[\protect\citename{Dickinson }1987]{dfd87}
Dickinson D.F., 1987, ApJ, 313, 408
\bibitem[\protect\citename{Dixon, Field \& Zare }1985]{dfz85}
Dixon R.N., Field D., Zare R.N., 1985, Chem. Phys. Lett., 122, 310
\bibitem[\protect\citename{Elitzur }1981]{el81}
Elitzur M., 1981, in `Physical Processes in Red Giants' 
 (I. Iben \& A. Renzini, eds.), Reidel, Dordrecht, p363
\bibitem[\protect\citename{Elitzur, Goldreich \& Scoville }1976]{egs76}
Elitzur M., Goldreich P., Scoville N., 1976, ApJ, 205, 384
\bibitem[\protect\citename{Field \& Gray }1988]{fg88}
Field D., Gray M.D., 1988, MNRAS, 234, 353
\bibitem[\protect\citename{Flower, Pineau des For\^{e}ts \& Walmsley }2006]{fpw06}
Flower D.R., Pineau des For\^{e}ts G., Walmsley C.M., 2006, A\&A, 449, 621
\bibitem[\protect\citename{Gaume \& Mutel }1987]{gm87}
Gaume R.A., Mutel R.L., 1987, ApJSS, 65, 193
\bibitem[\protect\citename{Gray }2001]{g01}
Gray M.D., 2001, MNRAS, 324, 57
\bibitem[\protect\citename{Gray, Doel \& Field }1991]{gdf91}
Gray M.D., Doel R.C., Field D., 1991, MNRAS, 252, 30
\bibitem[\protect\citename{Gray, Field \& Doel }1992]{gfd92}
Gray M.D., Field D., Doel R.C., 1992, A\&A, 262, 555
\bibitem[\protect\citename{Gray, Howe \& Lewis }2005]{ghl05}
Gray M.D., Howe, D.A., Lewis, B.M., 2005, MNRAS, 364, 783
\bibitem[\protect\citename{Jones et al. }1994]{jones94}
Jones K.N., Field D., Gray M.D., Walker R.N.F., 1994, A\&A, 288, 581
\bibitem[\protect\citename{Kylafis \& Norman }1990]{kn90}
Kylafis N.D., Norman C.A., 1990, ApJ, 350, 209
\bibitem[\protect\citename{Litvak }1969]{lit69}
Litvak M., 1969, ApJ, 156, 471
\bibitem[\protect\citename{Lucas }1980]{luc80}
Lucas R., 1980, A\&A, 84, 36
\bibitem[\protect\citename{Offer, van Hemmert \& van Dishoeck }1994]{off94}
Offer A.R., van Hemmert M.C., van Dishoeck E.F., 1994, J. Chem. Phys., 100, 362
\bibitem[\protect\citename{Pavlakis \& Kylafis }1996]{pk96}
Pavlakis K.G., Kylafis N.D., 1996, ApJ, 467, 309
\bibitem[\protect\citename{Piehler \& Kegel }1989]{pk89}
Piehler G., Kegel W.H., 1989, A\&A, 214, 339
\bibitem[\protect\citename{Randell et al. }1995]{ran95}
Randell J., Field D., Jones K.N., Yates J.A., Gray M.D., 1995, A\&A, 300, 659
\bibitem[\protect\citename{Rybicki \& Lightman }1979]{rl79}
Rybicki G.B., Lightman A.P., 1979, `Radiative Processes in Astrophysics', Wiley,
New York
\bibitem[\protect\citename{Scharmer \& Carlsson }1985]{sac85}
Scharmer G.B., Carlsson M., 1985, J. Comp. Phys. 59, 56
\bibitem[\protect\citename{Sobolev }1986]{sob86}
Sobolev A.M., 1986, Sov. Ast. 30, 399
\bibitem[\protect\citename{Sobolev }1989]{sob89}
Sobolev A.M., 1989, Astronomische Nachrichten, 310, 343
\bibitem[\protect\citename{Sobolev \& Deguchi }1994a]{sobdeg94}
Sobolev A.M., Deguchi S., 1994, ApJ, 433, 719
\bibitem[\protect\citename{Sobolev \& Deguchi }1994b]{sd94}
Sobolev A.M., Deguchi S., 1994, A\&A, 291, 569
\bibitem[\protect\citename{Stift }1992]{stift}
Stift M.J., 1992, Lecture Notes in Physics, 401, 431
\bibitem[\protect\citename{Yates Field \& Gray }1997]{yfg97}
Yates J.A., Field D., Gray M.D., 1997, MNRAS, 285, 303
\bibitem[\protect\citename{Yu }2005]{yu05}
Yu Z., 2005, Annals of Shanghai Observatory (English abstract), 26, 95

\end{thebibliography}
\end{document}